# Tree-based Visualization and Optimization for Image Collection

Xintong Han, Chongyang Zhang, Weiyao Lin, Mingliang Xu, Bin Sheng, and Tao Mei

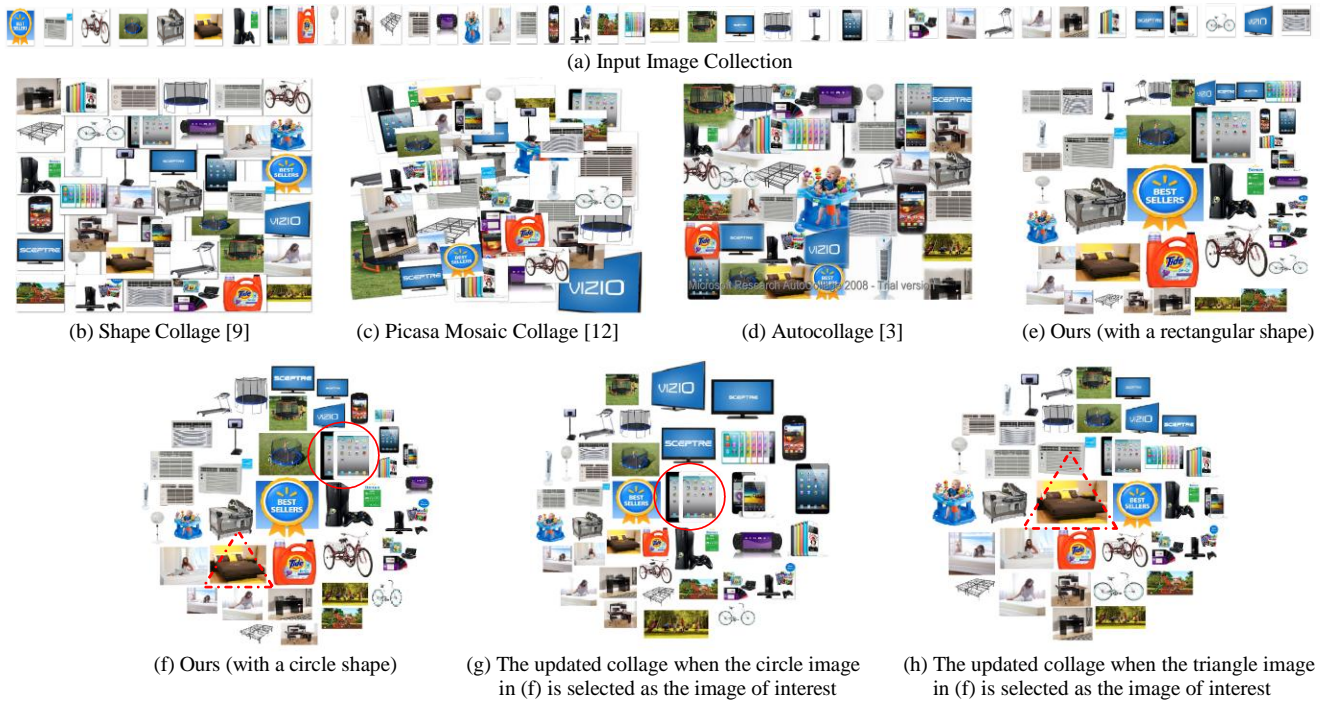

Figure 1. Image collection visualization comparison.

*Abstract*—The visualization of an image collection is the process of displaying a collection of images on a screen under some specific layout requirements. This paper focuses on an important problem that is not well addressed by the previous methods: visualizing image collections into arbitrary layout shapes while arranging images according to user-defined semantic or visual correlations (e.g., color or object category). To this end, we first propose a property-based tree construction scheme to organize images of a collection into a tree structure according to user-defined properties. In this way, images can be adaptively placed with the desired semantic or visual correlations in the final visualization layout. Then, we design a two-step visualization optimization scheme to further optimize image layouts. As a result, multiple layout effects including layout shape and image overlap ratio can be effectively controlled to guarantee a satisfactory visualization. Finally, we also propose a tree-transfer scheme such that visualization layouts can be adaptively changed when users select different "images of interest". We demonstrate the effectiveness of our proposed approach through the comparisons with state-of-the-art visualization techniques.

This work was supported in part by the National Science Foundation of China grants (61471235, 61472370, 61202207, and 61202154) and the Chinese national 973 grants (2013CB329603). Corresponding authors are Chongyang Zhang and Weiyao Lin.

X. Han, C. Zhang, and W. Lin are with the Department of Electronic Engineering, Shanghai Jiao Tong University, Shanghai 200240, China (email: {hxt_1123, sunny_zhang, wylin }@sjtu.edu.cn).
M. Xu is with the School of Information Engineering, Zhengzhou University, Zhengzhou 450001, China (email: iexumingliang@zzu.edu.cn).
B. Sheng is with the Department of Computer Science and Engineering, Shanghai Jiao Tong University, Shanghai 200240, China (email: shengbin@cs.sjtu.edu.cn).
T. Mei is with Microsoft Research, Beijing 100080, China (email: tmei@microsoft.com).

## I. Introduction

The visualization of an image collection is the process of displaying a collection of images on a screen under some specific layout requirements, such as layout shape [2, 3, 9, 11, 20, 31], image size [2, 4, 11], overlap among images [2, 6-8], and so on. With the rapid growth of digital image content, it is of increasing importance to effectively visualize the huge amount of image collections. For example, one wants to create a collage of vacation photos, summarize image search results of a given query, or design a poster including logos of the world's top 50 largest companies. Although many algorithms have been proposed for image collection visualization [1-14, 20-21], the following challenging problems still need to be addressed.

● **Visualizing into arbitrary layout shapes.** In many image visualization applications, people may often want images to be visualized into arbitrary shapes such that various layout effects can be created. However, most existing visualization techniques only focus on fixed layouts (e.g., rectangle) which cannot be extended to arbitrary shapes [3, 12, 14, 20, 21]. Although some methods [9, 19, 25, 26] can create various layout shapes, they still have limitations in avoiding the overlap among images and precisely controlling the correlations among images.

● **Properly arranging images according to user-defined correlations.** Since organizing images by their correlations is beneficial in many applications [27, 37-39], people may often want images in the visualization layout to be arranged according to their desired property correlations (e.g., color and category) [33-34], so that images with high correlation or similar properties can be placed closely. However, this functionality cannot be achieved in many visualization algorithms [2-4, 9-12, 29] due to the negligence of image correlations. Although some projection-based methods [5, 13, 28] can embed image correlations in a layout, their capabilities in creating arbitrary shapes and avoiding image overlaps are limited.

● **Adaptive updating layouts according to user-selected images of interest.** Moreover, people may also want visualization layouts to be able to adaptively change according to their interests (i.e., their selected images of interest). However, most existing visualization methods only focus on the visualization of initial layouts, while the update of layouts is neglected or not properly addressed.

Therefore, in this paper, we focus on addressing the above problems and propose a tree-based visualization approach. Figure 1 shows an example of our approach. Figure 1 (a) is a collection of input images of "the best-selling products in Walmart", (b)-(d) are the results from existing visualization techniques [3, 9, 12], while (e)-(f) are the results produced by our approach. In practice, people may want products of the same category (e.g., electronic products, and furniture) to be placed closely in a layout for efficient browsing. However, in (b)-(d), this requirement cannot be always satisfied, where the images are sometimes placed disorderly. In (e)-(f), users can flexibly define "product category" as their desired property and images in a layout can then be arranged accordingly. Furthermore, when users desire images to be displayed with different shapes, our approach is able to visualize this image collection into arbitrary layout shapes, as shown in Figure 1 (f). Finally, when users want to create "personalized" image collages such that images of their interested product categories are displayed more distinctively in the collage layout, our approach can also adaptively update the collage layout according to different user-selected "images of interest", as shown in Figure 1 (g)-(h).

*A. Related Work*

The problem of image collection visualization has been studied for years and a variety of approaches have been proposed [1-14, 20-21, 25-29, 36-38]. An extensive summarization about the existing techniques used in image collection visualization can be found in [33-34, 39]. However, due to the contradicting nature of creating arbitrary layout shape and maintaining image correlations (i.e., in order to precisely fit images into arbitrary layout shapes, it is more suitable to disregard the correlation among images and place images only according to the layout shape; while on the other hand, it is difficult to create arbitrary layout shapes when maintaining the image correlations), it is still a challenging problem to integrate both issues into the same framework.

Some algorithms visualize image collections according to predefined templates [1, 14, 20, 21, 37, 38]. For example, Kelly and Ma visualize images by a hierarchical radial focus template such that higher-ordered images are displayed on the inner ring of the template [14]. Strong and Gong introduce a self-sorting map to organize images into a structured layout [20]. Schoeffmann et al. [37, 38] arrange images using color-based similarity and assign them to a 3-D ring or a 3D global interface. However, since these methods utilize predefined templates for visualization, they cannot create arbitrarily shaped layouts.

Besides, many image-collage approaches aim to create image collages by visualizing images within a collage shape [2-4, 9-12, 29, 40-41]. For example, some saliency-based methods [2, 3, 40] create collage results by finding an optimal rectangular region for each image inside a collage shape while avoiding the occlusion of images' salient parts. Other mosaic-based methods [10, 41] aim to select suitable images as cut-outs and fit them in a pre-defined mosaic layout. Although many of these methods can flexibly create arbitrary layout shapes, images are often disorderly placed in their collage or mosaic results due to the negligence of image correlations.

Moreover, projection-based visualization algorithms are also widely used for image collection visualization [5-8, 13, 28, 30, 36]. These methods first calculate correlations among images in a high-dimensional feature space. Then, images are projected into a visualization space by some dimensionality reduction techniques [5, 13, 28, 36, 46-48]. Since directly projecting images often results in severe image-wise overlap or occlusion in a visualization layout, other works are further proposed to reduce this overlapping problem by utilizing template guidance [6, 36] or graphical optimization [7, 8]. Although these projection- based methods can embed the image correlations in the layout, most of these methods do not have the capability to create arbitrary shaped layouts. Besides, many of them still have overlap problems when the number of images increases.

Furthermore, in recent years, some researchers [25, 26] proposed to utilize a Centroidal Voronoi Tessellation (CVT) to balance and optimize an initial image distribution, such that both image correlation maintenance and arbitrary layout creation can be achieved. However, they still have the following limitations: (1) Although these CVT-based methods can properly avoid image overlaps in some layout results, they may still have obvious overlaps under complex layout shapes or densely-overlapped initial image distributions. (2) These methods model image properties in a parallel way. This makes them difficult to handle the multi-layer property correlation among images (e.g., first grouping images according to "color" property and then within each image group of similar color, further putting together images with similar "texture" property). Although Reinert et al. [25] propose to weight the relative importance among properties according to users' manipulation on primitive images, it still has limitations in creating precise layouts according to user-desired multi-layer correlations.

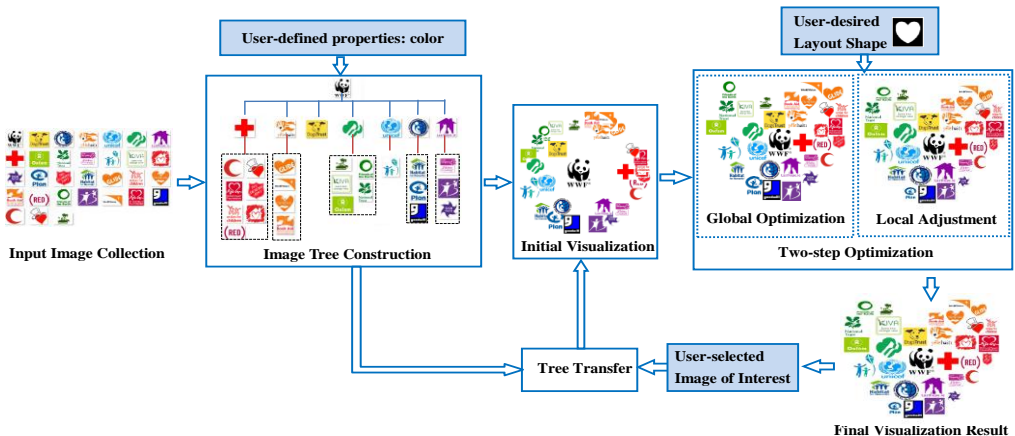

Figure 2. The overview of the proposed approach. (Best viewed in color)

Finally, most of the existing image collection visualization methods only focus on the suitable visualization of initial layouts while the update of visualization layouts is not properly addressed [5, 13, 22] or even inapplicable [1-4, 6-12] when user select different "images of interest". This makes them difficult to adaptively create "personalized" visualization layouts. Although some methods allow layout update through human interaction [25, 26, 36], these methods still have limitations in visualizing all images in a collection or maintaining the original image correlations in an updated layout.

*B. Our Approach*

To address the above problems, we propose a tree-based visualization approach. Our proposed approach is a hybrid of the projection-based method and the image collage method. First, we organize images of a collection into a tree structure to encode user-desired correlations or properties for images. Then, by projecting this tree structure into a visualization plane, the desired correlations among images can be well reflected in the layout. Furthermore, we propose a two-step visualization optimization scheme to optimize image layouts such that layout effects including image overlap and layout shape can be properly controlled in the final layout while maintaining the original image correlation at the same time. Finally, we also propose a tree-transfer scheme which enables the adaptive updating of visualization layouts according to different user-selected "images of interest".

The rest of the paper is organized as follows. Section II describes overview of the proposed approach. Section III-V present the proposed tree-construction scheme, the two-step visualization optimization scheme, and the tree transfer scheme, respectively. Section VI shows the experimental results and Section VII concludes the paper.

## II. OVERVIEW OF THE APPROACH

The overview of our proposed approach is shown in Figure 2. Our approach mainly includes four components. First, the "image tree construction" component is used to organize images into a tree structure such that user-desired visual or semantic correlation among images can be well embedded. After that, we need to project this tree into a visualization space to create a visualization layout. Since simple projection cannot effectively create arbitrary layout shapes or avoid image overlaps, our approach introduces two components to transfer a tree into a final visualization layout: first, projection methods are applied to project a tree to create an initial visualization layout (i.e., the "initial visualization" component) [5, 13]. Then, the "two-step optimization" component further updates and optimizes the initial layout through the "global optimization" and "local adjustment" steps. Since projection methods can reflect correlations of a tree in the layout space while "two-step optimization" can effectively control layout effects, by combining these components, both the desired image correlations and the layout effects can be guaranteed in the final layout. Finally, if a user selects his/her "image of interest", the "tree transfer" component will transfer the tree structure accordingly. And this transferred tree will then go through the "initial visualization" and the "two-step optimization" components to achieve the updated layout.

In the following, we will discuss the four components in detail. Note that "*image tree construction*" and "*two-step optimization*" are the *key contributions* of the paper.

## III. TREE CONSTRUCTION AND INITIAL VISUALIZATION

*A. Image Tree Construction*

As mentioned, the "image tree construction" component aims to organize images according to user-defined properties such that images with high property correlations are placed close to each other in a layout. Inspired by the previous studies on radius and hyperbolic visualization [13, 14], we observe that hierarchical (or tree) structures are a suitable way to model the multi-layer visual or semantic property correlations in an image collection. This is because:

(1) A tree structure can flexibly model multi-layer property correlations among images. For example, in an image search result collection of a query "apple" in Figure 3, we may first like images of the same "object types" (i.e., fruits, electronic device, logo) to be placed together. Then, within each object type, we may further want images with similar "color" to be placed close to each other. Thus, by using a tree structure, both properties can be well modeled by using one layer to model the "object type" property and another layer to model the "color" property, as Figure 3 (b) shows.

(2) A tree structure can be easily combined with projection methods [5, 13] to create proper initial layouts. Since many projection methods [5-8, 13] can create layouts from a tree structure which projects sibling images of a tree close to each other, by combining a tree structure with projection methods, user-desired properties can be effectively reflected in a visualization layout (such as Figure 3 (c)).

Therefore, in this paper, we propose a property-based tree construction (PTC) scheme to organize images into a tree according to user-desired properties.

Figures 3-4 show an example of using our scheme to construct an image tree. In Figure 3 (a), we have an image collection which is arranged in a 1-D array. A subset of the image collection and its features are shown in Figure 4 (a). In Figure 4 (a), two features are defined to model image property correlations: the semantic "objects type" feature and the visual "color histogram" feature. We call these features as the user-defined "properties". As a result, every image $I_i$ will have a property vector $V_i=[P_{i,1}, P_{i,2}]$, where $P_{i,1}$ is image $I_i$'s semantic "object type" property feature (e.g., fruit, logo) and $P_{i,2}$ is the visual color histogram of the image. The target of our PTC scheme is to organize these images into a tree according to these properties (i.e., first place images of the same "object type" together (in the same branch) and then within each branch, place images with similar "color histogram" together (in the same sub-branch), as Figure 3 (b)).

Algorithm 1 and Figure 4 (b)-(f) show the detailed process of using our tree construction scheme. From Algorithm 1 and Figures 4 (b)-(f), we can see that in the PTC scheme, images in a collection are first arranged in a 1-D array. Then, one image is selected to be the root image $I_{root}$ and the other images are added sequentially into a tree according to their order in the 1-D array.

When adding image $I_i$, the dissimilarity between $I_i$ and images in the second level (i.e., the level next to the root) are first calculated to check the suitableness of inserting $I_i$ into this level. Note that this dissimilarity measures the correlation of the desired property between images. If there are no similar images in this level, $I_i$ should be placed in this level to be the ancestor node of a new branch for a new property content. For example, in Figure 4 (d), since the property of the cellphone image $I_3$ is "electronic device", it is different from the "fruit" property of the apple image $I_2$. Thus, it is placed at the second level to start a branch. On the contrary, if there are similar images to $I_i$ in the second level, in order to guarantee correlated images being organized in the same branch, $I_i$ should be moved to the third level as the child of $I_i$'s most similar image in the second level. For example, in Figure 4 (e), since image $I_4$ has the same "fruit" property to $I_2$ in the second level (left figure in (e)), $I_4$ is moved to the third level to be the child of $I_2$ (right figure in Figure 4 (e)).

Several issues need to be mentioned on the PTC scheme:

(1) Note that different properties can be used in different levels in a tree when calculating the dissimilarity among images. In this way, our PTC scheme can effectively handle multiple property correlations in a visualization layout. For example, with the tree structure of Figure 3 (b), images can be organized such that images containing the same object type (i.e., fruit, logo, electronic) are grouped together while within each group, images with similar colors are further grouped together (as shown in Figure 4 (c)).

(2) When users define different sets of properties (e.g., switch the order of the properties or define new properties), the PTC scheme can flexibly create different image trees accordingly. For example, Figure 3 (e) is another image tree constructed by our PTC scheme when switching the order of the two properties in Figure 4 (a) (i.e., users first want to place images of similar color together and then within each color group, place images of the same object type together). And Figures 3 (d) and (f) show the final visualization results of our approach for the image trees in (b) and (e). From Figures 3 (d)-(f), we can see that our PTC scheme can flexibly organize images according to different user-defined properties, thus making the final visualization results fulfill different user requirements.

(3) The property content for each image can be either achieved by manual tagging or automatically parsed by their associated tags (e.g., from Flickr) or through feature extraction or image annotation algorithms [15-16, 32, 44-45]. In our experiments, we extract color histogram (for color property) and histogram of gradients (for texture property) [15] as the visual features for all images, and extract 3-4 semantic features for each image collection (each semantic feature represents one property and different semantic features are defined for different image collections). For image collections accompanied with tags, we parse tags which are related to the defined properties and use them as images' semantic features. For image collections without tags, we manually tag the semantic features for each image. Overall, a total of 5-6 features are extracted for each image collection, and users can select a subset of these features to build an image tree and create the corresponding visualization result. Moreover,

(4) The root image of a tree can be selected either manually or by some automatic methods [15-16]. Also, the 1-D array can be arranged randomly or by some automatic methods [17, 24, 32]. In our experiments, we simply select the image which shares the least common properties with the other ones as the root and organize the input 1-D image array by ranking them according to user-defined properties [17, 24].

(5) In our experiments, the dissimilarity measure $d_f(\cdot)$ in Algorithm 1 is calculated by the following ways: The dissimilarity of two visual features (e.g., color histogram) is calculated by 1-$HI(P_{i,f}, P_{k,f})$, where $P_{i,f}$ and $P_{k,f}$ are the visual features describing the $f$-th property in images $I_i$ and $I_k$ and $HI(\cdot)$ is the histogram intersection similarity [35]. The dissimilarity of two semantic features is calculated by $NE(P_{i,g}, P_{k,g})$ where $P_{i,g}$ and $P_{k,g}$ are the tag values describing the $g$-th property for images $I_i$ and $I_k$. $NE(\cdot)=0$ if $P_{i,g}=P_{k,g}$, and 1 otherwise.

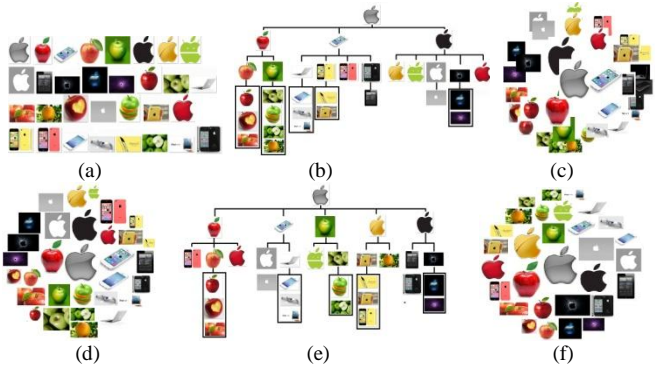

Figure 3. An example of our PTC scheme. (a) The input image search result collection of the query "apple". (b) The tree structure constructed for (a) using the properties in Figure 4 (a). (c) The initial visualization result created from (b) by hyperbolic projection. (d) The final visualization result for the image tree in (b). (e) Another image tree constructed for (a) when switching the two properties in Figure 4 (a). (f) The final visualization result for the image tree in (e). (Best viewed in color)

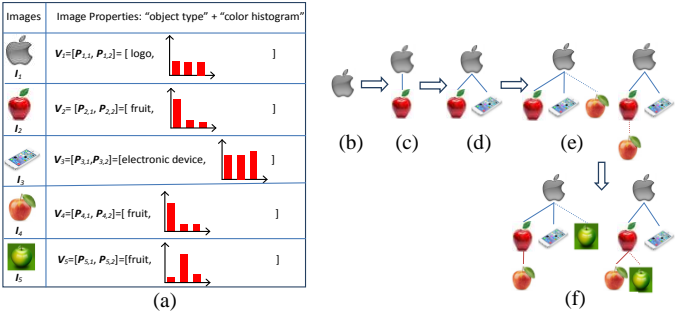

Figure 4. An example describing the detailed process of the tree construction scheme. (a): The first column of the table is the input image collection. The second column is the properties: the semantic "objects type" feature and the visual "color histogram" are used to describe the properties of every image (note that each image $I_i$ is represented by a property vector $V_i=[P_{i,1}, P_{i,2}]$, where $P_{i,1}$ is image $I_i$'s semantic "object type" property feature (e.g., fruit, logo) and $P_{i,2}$ is the visual color histogram feature of the image). (b)-(f): The images are added sequentially into the tree.

### B. Initial Visualization

After images are organized into a tree, the "initial visualization" component will project this tree into a visualization space to create an initial layout.

In this paper, hyperbolic projection [5, 13] is utilized which projects an image tree into a hyperbolic space to create an initialized visualization layout. We choose hyperbolic projection because: (1) it can guarantee high-level images being displayed more distinctively in a layout (i.e., images closer to the root node in a tree will be displayed closer to the layout center with larger sizes). (2) It can also properly maintain the original image correlations in a tree structure (i.e., images in the same branch are placed closely after projection). It should be noted that the framework of our approach is general. In practice, other initialization methods can also be utilized in our approach as long as they can properly initialize images' locations and sizes in a layout.

Figures 3 (c) and 5 (c) show two initial visualization results created by hyperbolic projection. From these figures, we can see that by hyperbolic projection, higher-level images are displayed more distinctively. For example, in Figure 5 (c), images of 2013 Oscar final winners (i.e., images in the second level of the tree in Figure 5 (b)) are displayed more distinctively than images of nominees (i.e., images in the third level of the tree). Besides, correlated images are displayed closely (e.g., the red apple images are located close to each other in Figure 3 (c)). However, although tree hierarchies and image correlations are effectively maintained in an initial visualization layout, it still suffers from the problems of severe overlap among images as well as failure to fit inside a user-defined target layout shape. Thus, a two-step visualization optimization (TSVO) scheme is proposed to further optimize the layout effect.

---

**Algorithm 1 Image Tree Construction**

**Input:** $N$ images $I_1,...,I_N$ and their corresponding property vectors $V_1,...,V_N$, where $V_i=[P_{i,1}, P_{i,2}, ..., P_{i,F}]$, and $P_{i,f}$ is the $f$-th property of image $I_i$ and $F$ is the total number of properties to describe each image.
**Output:** the organized image tree including $N$ images: $Tree=\{(I_i, I_{Qi}, J_i) \mid i=1,...,N\}$ where $I_{Qi}$ is the parent node of $I_i$ and $J_i$ is the level number of $I_i$ in the tree. And the maximum level of the tree is $J_{max}=F+1$ where $F$ is the total number of properties for an image.
**Algorithm:**
1. Put the $N$ images into a 1-D array
2. Set ***Tree*** as an empty set
3. Choose one image as the tree root $I_{root}$ and set $I_{Qroot}=I_{root}$, $J_{root}=0$
4. Add ($I_{root}$, $I_{Qroot}$, $J_{root}$) into ***Tree***
5. for $i = 1:N$ and $I_i \neq I_{root}$    //sequentially add each image into the ***tree***
6.    $I_{Qi}=I_{root}$, $J_i =1, f=1$;   // initialize the parent node and level of $I_i$
7.    while $J_i <= J_{max}$-1
8.       $idc_{finish}=1$, $d_{min}=\infty$;   // initialize the finish indictor $idc_{finish}$ and $d_{min}$
9.       for $I_k \in$ ***Tree***, $J_k= J_i$, and $I_{Qk}= I_{Qi}$
10.         if $d_f(P_{i,f}, P_{k,f})<T_f$ and $d_f(P_{i,f}, P_{k,f})< d_{min}$   //$T_f$ is a threshold and //$d_f(P_{i,f}, P_{k,f})$ is the dissimilarity measure
11.           $idc_{finish}=0$; $I_{Qi} = I_k$; $d_{min}= d_f(P_{i,f}, P_{k,f})$;
12.         end if
13.       end for
14.       if $idc_{finish} =0$   //find image with similar property $P_{i,f}$
15.         $f=f+1$; $J_i = J_i +1$;   // check the next level and the next property
16.       else
17.         break;   // the current level is the right level for $I_i$
18.       end if
19.    end while
20.    Add ($I_i$, $I_{Qi}$, $J_i$) into ***Tree***
21. end for

---

### IV. TWO-STEP VISUALIZATION OPTIMIZATION

The proposed TSVO scheme formulates constraints (e.g., layout shape, overlap, and image size) as a series of costs and utilizes a two-step way to optimize these costs for achieving optimized layout. Some results after the two optimization steps are shown in Figures 2 and 5 (d)-(e).

It should be noted that although some image collage algorithms [2-4, 9-13] also create collage layouts by optimizing some energy functions, our TSVO scheme is different from them in: (1) When optimizing a layout, our TSVO scheme also maintains the original property correlation in an image tree. Comparatively, most image collage algorithms do not consider keeping the correlation among images. (2) After the global optimization step, our TSVO scheme also introduces a local optimization step to locally update images, while most image collage algorithms only perform a global optimization process.

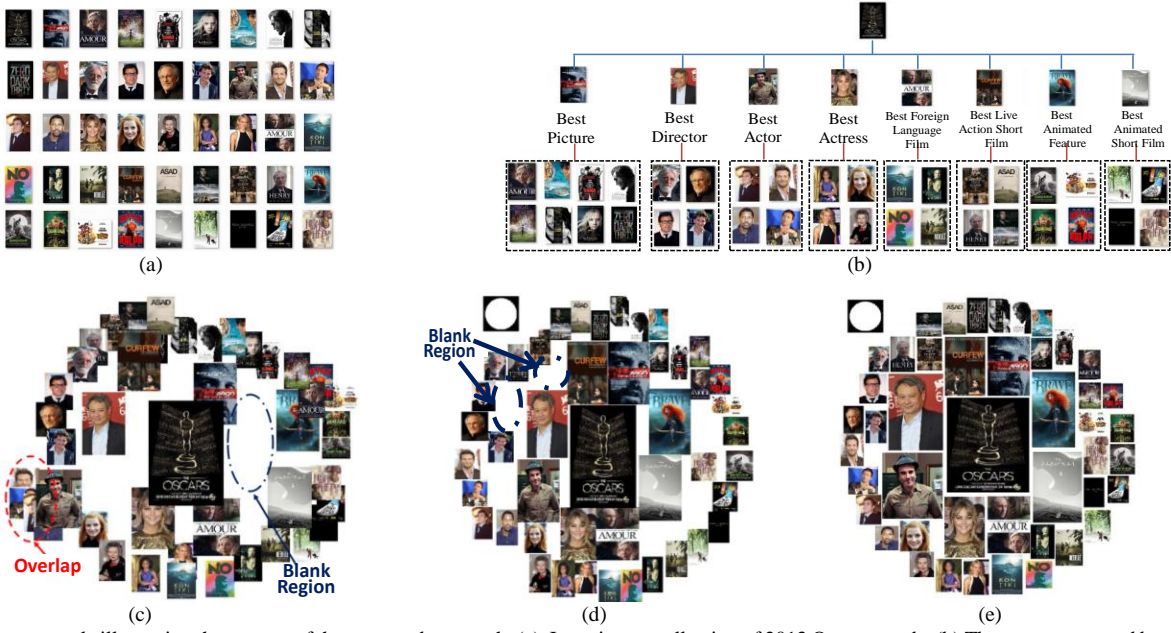

Figure 5. Another example illustrating the process of the proposed approach. (a) Input image collection of 2013 Oscar awards. (b) The tree constructed based on the "award category" property of the images (e.g., best director, best movie, best animated feature). (c) Initial visualization result. (d) Result after global optimization. (e) Final result after the local adjustment step (the circle on the up-left corner in (d)-(e) is the target shape).

### A. Problem Formulation

Assuming that there are $N$ images $\{I_1, I_2, \ldots, I_N\}$ that have been initially visualized by the "initial visualization" component, we introduce the following items that should be optimized in TSVO scheme:

● **Preserving property correlations in a tree.** Since we want image property correlations in a tree structure to be maintained in the final layout, the TSVO scheme should be able to preserve these correlations during the layout update process. Therefore, we introduce a correlation preservation cost to handle this issue as:

$$E_t(\mathbf{L}) = \frac{1}{N-1}\sum_{i=2}^{N}\left(1 - exp\left(-\frac{d(I_i, I_{Q_i})}{\sigma_i}\right)\right) \quad (1)$$

where $\mathbf{L}=\{(x_1, y_1), (x_2, y_2), \cdots, (x_N, y_N)\}$ is the set for all $N$ image locations in a layout where $(x_i, y_i)$ are the coordinate of the center point of image $I_i$ in the layout. $d(I_i, I_{Q_i})$ is the Euclidean distance between image $I_i$ and its parent image $I_{Q_i}$ in the layout. $\sigma_i = -\sqrt{w_{Q_i}^2 + h_{Q_i}^2}/ln(1-T)$ is a curvature-rolling factor where $w_{Q_i}$ and $h_{Q_i}$ are the weight and height for image $I_{Q_i}$ in the layout and $T$ is set to be 0.5 in our paper. By minimizing the structure preservation cost in Eqn. (1), children images in a tree can be placed close to their parents in the final layout.

● **Preserving hierarchical structures in a tree.** Besides the property correlations, the hierarchical structure of a tree should also be maintained in a layout (i.e., high-level images should be displayed more distinctively). Therefore, a level preservation cost is also introduced by:

$$E_l(\mathbf{L}) = \frac{1}{N-1}\sum_{i=2}^{N}\left(1 - exp\left(-\frac{|d(I_i, I_{root}) - R(I_i)|}{\rho_i}\right)\right) \quad (2)$$

where $d(I_i, I_{root})$ is the Euclidean distance between image $I_i$ and the root image $I_{root}$ in a tree structure. $R(I_i)$ is the radius of image $I_i$ which controls the distance between $I_i$ and the root image in a visualization layout. $R(I_i)$ is calculated by $R(I_i) = R_{\mathbf{G}_j}$ for $I_i \in \mathbf{G}_j$, where $\mathbf{G}_j$ is the group of images in $j$-th level of an image tree. $R_{\mathbf{G}_j} = R_{\mathbf{G}_{j-1}} + \sum_{I_k \in \mathbf{G}_{j-1}} \sqrt{(w_k^o)^2 + (h_k^o)^2}/N_{\mathbf{G}_{j-1}}$, where $R_{\mathbf{G}_{j-1}}$ is the radius for image group $\mathbf{G}_{j-1}$ in $j$-1'th level in the tree (i.e., $\mathbf{G}_j$'s parent level) and the radius for the root image $R_{\mathbf{G}_0}$ is set as 0. $w_k^o$ and $h_k^o$ are the width and height of image $I_k$ in the initial visualization layout (i.e., layout from the "initial visualization" component). From the above equation, we can see that images in $j$-th level are given the same radius $R_{\mathbf{G}_j}$ where $R_{\mathbf{G}_j}$ is calculated by the radius of $\mathbf{G}_j$'s parent level $R_{\mathbf{G}_{j-1}}$ plus the average diagonal length for images in $\mathbf{G}_j$'s parent level. Besides, $\rho_i = -\sqrt{w_i^2 + h_i^2}/ln(1-T)$ in Eqn. (2) is another curvature-rolling factor where $w_i$ and $h_i$ are the weight and height for image $I_i$ in the layout and $T$=0.5 in our paper. By Eqn. (2), we can control the distance among tree levels and avoid low-level images from being mixed with high-level images.

● **Image size.** Image size is an important issue affecting final visualization results. Firstly, it is desirable that image sizes can be controlled such that they will not be displayed extremely large or small (i.e., image sizes in a layout can be controlled within a specific range). Secondly, since people may often desire that the images can be large to be clearly seen, we also want image sizes to be as large as possible within the size ranges (i.e., close to the upper bound of the size range). Therefore, we define an image size cost as well as an image size constraint as shown in Eqns (3) and (4), respectively.

$$E_s(\mathbf{R}) = \frac{1}{N}\sum_{i=1}^{N} \frac{w_i^u h_i^u - w_i h_i}{w_i^u h_i^u - w_i^l h_i^l} \quad (3)$$

$$w_i^l \leq w_i \leq w_i^u, h_i^l \leq h_i \leq h_i^u, \quad (4)$$

where $E_s(\mathbf{R})$ in Eqn. (3) is the image size cost. $\mathbf{R}=\{(w_1, h_1), (w_2, h_2), \cdots, (w_N, h_N)\}$ is the set for all $N$ image sizes in a layout where $w_i$ and $h_i$ are the weight and height for image $I_i$ in the layout. $w_i^u$ and $h_i^u$ are image $I_i$'s upper size bound while $w_i^l$ and $h_i^l$ are image $I_i$'s lower size bound. If image $I_i$ is displayed in its minimum size (i.e., $w_i^l h_i^l$), its size cost will be 1. If its size reaches the upper bound, its size cost is 0. Thus by minimizing this cost, the images can be displayed of larger sizes. Furthermore, the image size constraint in Eqn. (4) is defined to guarantee that images can be displayed within their size ranges. In this paper, the size ranges are set by default as $w_i^l = 0.8w_i^o$, $w_i^u = 1.2w_i^o$, $h_i^l = 0.8h_i^o$, $h_i^u = 1.2h_i^o$ where $w_i^o$ and $h_i^o$ are the size of image $I_i$ in the initial visualization layout (i.e, $w_i^o$ and $h_i^o$ are the width and height of image $I_i$ after the "initial visualization" component, as shown in Figs 3 (c) and 5 (c)).

● **Overlap among images.** As mentioned, overlap among images is another key issue that should be properly avoided or controlled. Thus, we define the overlap constraint as:

$$\sum_{q=1,q\neq i}^{N} \frac{o_{i,q}}{w_i h_i} \leq T_o \quad (5)$$

where $o_{i,q}$ is the overlap size between images $I_i$ and $I_q$ in a layout and $T_o$ is the maximum allowed overlap threshold. $T_o$ can be defined by users to control the level of maximum overlap in the final layout. In this paper, $T_o$ is set to be 0 meaning that no overlap is allowed in the layout.

● **Layout shape.** Finally, in order to visualize images within arbitrary layout shapes, we also introduce a layout shape constraint as shown in Eqn. (6).

$$(x_i, y_i) \in \mathbf{SP} \text{ for } i = 1, \ldots, N \quad (6)$$

where $(x_i, y_i)$ is the coordinate of the center point of image $I_i$ in a layout and $\mathbf{SP}$ is the user-defined layout shape.

To handle the above items, we model the following optimization problem to create an optimized layout.

$$(\mathbf{L}^{opt}, \mathbf{R}^{opt}) = arg \min_{\mathbf{L},\mathbf{R}}(\omega_t E_t(\mathbf{L}) + \omega_l E_l(\mathbf{L}) + \omega_s E_s(\mathbf{R})) \quad (7)$$

s.t. $(x_i, y_i) \in \mathbf{SP}$,
$w_i^l \leq w_i \leq w_i^u, h_i^l \leq h_i \leq h_i^u$,
$\sum_{q=1,q\neq i}^{N} \frac{o_{i,q}}{w_i h_i} \leq T_o$, for $i = 1, \ldots, N$

where $\mathbf{L}^{opt}$ and $\mathbf{R}^{opt}$ are the final optimized sets for all $N$ images' locations and sizes in a layout. $\omega_t$, $\omega_l$, and $\omega_s$ ($0\leq\omega\leq1$, $\omega_{tree}+\omega_{level}+\omega_{size}=1$) are the weighting parameters to balance the relative importance of the level preservation cost, the tree correlation preserving cost, and the image size cost. In Eqn. (7), the first two costs $E_t(\mathbf{L})$ and $E_l(\mathbf{L})$ aim to maintain the original tree structure in a layout while the image size cost $E_s(\mathbf{R})$ and the three constraints control the other layout effects including overlap and layout shape. Thus, satisfactory layout results can be achieved.

However, the optimization in Eqn. (7) is difficult to solve due to the non-analytical and complicated constraints. Therefore, we further propose to solve Eqn. (7) in a two-step way. That is, a global optimization step is first applied to achieve a roughly optimized layout. Then a local adjustment step is applied to locally adjust images to fulfill all layout constraints.

### B. Global Optimization

The global optimization step simplifies Eqn. (7) to achieve a roughly optimized layout. It is modeled by:

$$(\mathbf{L}_j^{opt}, \mathbf{R}_j^{opt}) = arg \min_{\mathbf{L}_j,\mathbf{R}_j} \begin{pmatrix} \omega_t E_t(\mathbf{L}_j) + \omega_l E_l(\mathbf{L}_j) + \omega_s E_s(\mathbf{R}_j) \\ + \omega_o E_o(\mathbf{L}_j, \mathbf{R}_j) + \omega_{sp} E_{sp}(\mathbf{L}_j) \end{pmatrix} \quad (8)$$

s.t. $w_i^l \leq w_i \leq w_i^u, h_i^l \leq h_i \leq h_i^u, I_i \in \mathbf{G}_j$

where $\mathbf{L}_j$ and $\mathbf{R}_j$ are the sets of locations and sizes for images in $j$-th level in a tree and $\mathbf{L}_j^{opt}$ and $\mathbf{R}_j^{opt}$ are the optimized results. $\mathbf{G}_j$ is the image set for $j$-th level. $w_i^l$, $w_i^u$, $h_i^l$, $h_i^u$ are the lower and upper bounds for image $I_i$'s width and height. $E_t(\mathbf{L}_j)$, $E_l(\mathbf{L}_j)$, and $E_s(\mathbf{R}_j)$ are the same costs as in Eqn. (7). $\omega_t$, $\omega_l$, $\omega_s$, $\omega_o$, and $\omega_{sp}$ are the weighting parameters to balance the cost importance. $E_o(\mathbf{L}_j, \mathbf{R}_j)$ is the overlap cost to penalize undesirable overlaps among images and $E_{sp}(\mathbf{L}_j)$ is the shape cost to measure whether images in $j$-th level are displayed inside a user-defined layout shape region. $E_o(\mathbf{L}_j, \mathbf{R}_j)$ and $E_{sp}(\mathbf{L}_j)$ are defined by:

$$E_o(\mathbf{L}_j, \mathbf{R}_j) = \frac{1}{N_{\mathbf{G}_j}} \sum_{I_i \in \mathbf{G}_j} \max \left( \sum_{I_q \in \mathbf{G}_{j-1}, q\neq i} \frac{o_{i,q}}{w_i h_i} - T_o, 0 \right) \quad (9)$$

$$E_{sp}(\mathbf{L}_j) = \frac{1}{N_{\mathbf{G}_j}} \sum_{I_i \in \mathbf{G}_j} \Delta(I_i) \left( 1 - exp\left(-\frac{d_m(I_i, \mathbf{SP})}{\rho_i}\right) \right) \quad (10)$$

where $o_{i,q}$ is the overlap between images $I_i$ and $I_q$ in a layout and $T_o$ is the maximum allowed overlap threshold between images. $\mathbf{G}_j$ is the group of images in $j$-th level of an image tree and $N_{\mathbf{G}_j}$ is the number of images in $\mathbf{G}_j$. $\mathbf{SP}$ is the area of a user-defined layout shape. $d_m(I_i, \mathbf{SP})$ is the Euclidean distance from $I_i$ to its nearest location inside the shape region $\mathbf{SP}$. $\Delta(I_i)$ is an indicator on whether $I_i$ is inside $\mathbf{SP}$ or not. $\Delta(I_i)=0$ if $I_i$ is inside $\mathbf{SP}$ and $\Delta(I_i)=1$ otherwise. $\rho_i$ is the same as in Eqn. (2).

From Eqn. (8), we can see that:

(1) The global optimization step mainly simplifies the optimization problem of Eqn. (7) by: (a) moving the overlap and shape region constraints in Eqn. (7) into the cost function (i.e., $E_o(\mathbf{L}_j, \mathbf{R}_j)$ and $E_{sp}(\mathbf{L}_j)$) such that a rough optimized layout can be easily achieved. (b) Performing optimization in a level-by-level way. That is, when optimizing the layout of $j$-th level images, only images in the current level and its higher levels are considered.

(2) Note that $\omega_o$ and $\omega_{sp}$ should be set obviously larger (e.g.,

100 times larger) than the other weighting factors in order to guarantee that the overlap and shape region constraints are properly addressed during optimization.

The simplified optimization problem in Eqn. (8) can be solved by the active-set algorithm [18]. Some results of the global optimization step are shown in Figures 2 and 5 (d).

*C. Local Adjustment*

From Figure 5 (d), we can see that the global optimization step still doesn't fully satisfy all the constraints in Eqn. (7) in: (1) Some images are still outside user-defined shape regions; (2) The overlap among some images still conflicts with the constraint; (3) The layouts are still not perfectly optimized with some blank regions not fully utilized. Thus, we further propose a local adjustment step to make a layout fully satisfy the constraints. This local adjustment step contains the following three major sub-steps:

● **Outside Image Moving** This sub-step moves images to guarantee that all images are located inside the target shape region SP. It can be described by:

(1) For all images that are out of shape region **SP**, we move them to their nearest locations inside **SP**.

(2) Re-solve Eqn. (8) to adjust the overlap and the layout of images. Note that since most image positions after global optimization will be inside **SP**, only few images will be updated and re-solving the minimization problem is fast.

(3) Repeat (1) and (2) until all images are inside **SP** or it reaches the maximum iteration time.

● **Image Size Updating** Since some images may still slightly conflict with the overlap constraints after global optimization, we further scale or update image sizes to fulfill the overlap requirements among all images [23].

● **Image Local Tuning** Finally, we further apply a local tuning sub-step to update image locations and sizes such that the blank regions inside a shape region can be more efficiently used by images. This sub-step is realized by solving the maximization problem as:

$$(x_i^*, y_i^*, w_i^*, h_i^*) = \arg\max_{\hat{x}_i, \hat{y}_i, \hat{w}_i, \hat{h}_i} (\hat{w}_i \hat{h}_i) \quad (11)$$

s.t. $\frac{2}{3} \leq \frac{\hat{h}_i}{\hat{w}_i} / \frac{h_i}{w_i} \leq \frac{3}{2}$, $\hat{w}_i \leq w_i^u$, $\hat{h}_i \leq h_i^u$, $\sum_k M_{\hat{I}_i}(k) \leq \sum_k M_{I_i}(k)$

$x_i - T_{range} \leq \hat{x}_i \leq x_i + T_{range}$, $y_i - T_{range} \leq \hat{y}_i \leq y_i + T_{range}$

where $h_i$, $w_i$, $(x_i, y_i)$ are the height, width, and location of image $I_i$ after the "image size updating" sub-step. $\hat{w}_i$, $\hat{h}_i$, $(\hat{x}_i, \hat{y}_i)$ are the width, height, location of a local tuned image $\hat{I}_i$ ($\hat{I}_i$ is achieved by tuning the size and location of $I_i$). $h_i^*$, $w_i^*$, $(x_i^*, y_i^*)$ are the optimized result. $T_{range}$ is a threshold that limits the local tuning range of $I_i$. $\sum_k M_{\hat{I}_i}(k)$ and $\sum_k M_{I_i}(k)$ are indicators to indicate whether pixel $k$ in image $I_i$ and $\hat{I}_i$ is overlapped with the other images or out of the shape region **SP**, respectively. $M_{I_i}(k)$ will be 1 if pixel $k$ of image $I_i$ overlaps with other images or out of **SP**, and 0 otherwise.

The local tuning in Eqn. (11) tries to maximize $I_i$'s size by searching a local region around its original location where the first constraint prevents the width/height ratio of the tuned image $\hat{I}_i$ from being too different from $I_i$. The second and third constraints limit the maximum size of $I_i$ to prevent images of lower levels being more distinct than the ones in the higher levels. The third constraint guarantees that the tuning process is restricted on the unused blank regions rather than the occupied regions. The last two constraints forbid images from moving too far away from their global optimization positions so as to maintain the overall image distribution. And Eqn. (11) can be effectively solved by dynamic programming [18]. In our experiment, we perform local tuning in a level-by-level way which sequentially tunes the images from higher levels to lower levels.

V. TREE TRANSFER

As mentioned, when users select their "images of interest", the layout should be able to adaptively change to create "personalized" layouts. Therefore, we propose a tree-transfer scheme to update an image tree according to the change of focus images.

If a user selects image $I_i$ as his/her image of interest (e.g., the image in the dashed rectangular in Figure 6 (a)), the tree will be transferred such that $I_i$ will become the root of the transferred tree while $I_i$'s parent image and sibling images adjacent to $I_i$ will become $I_i$'s child images and other parent-child relationships in the tree remain the same, as shown in Figure 6 (b). Once we have the transferred tree structure, hyperbolic projection is applied again to perform initial visualization of this new image tree. Finally, our two-step optimization scheme is performed to create the optimized layout, as shown in Figure 6 (c).

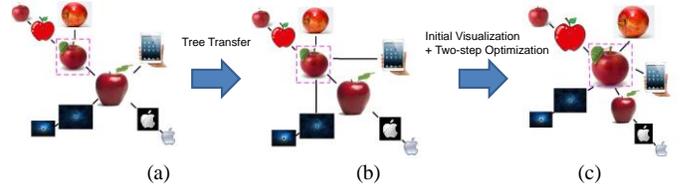

(a)  (b)  (c)

Figure 6. The process of tree transfer. (a) The original image tree. (b) The transferred tree when selecting the image in the dashed rectangle as the "focus image" by using this image as the root of the tree and using its parent and adjacent siblings as its children. (c) Initial visualization and two-step optimization are then used to create the updated visualization result for the selected image of interest.

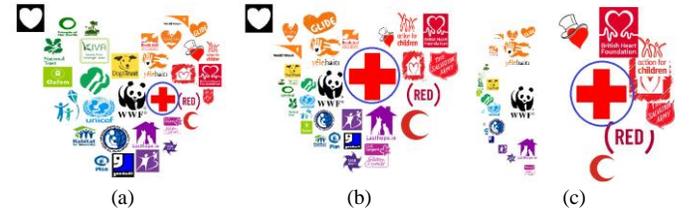

(a)  (b)  (c)

Figure 7. (a) The original layout in Figure 2. (b) The updated layout after tree transfer scheme when selecting the circled image as the "focus image". (c) The updated layout by [13] and [22].

As a result, images close to the "image of interest" will be displayed more distinctively while the correlations among images are well preserved. It should be noted that when selecting leaf images as images of interest, the transferred tree structures may be extremely unbalanced which may lead to unsatisfactory visualization results. Therefore, we further introduce a detector to check the balanceness of a transferred tree by evaluating the number of images in different tree branches. If a transferred tree is detected as extremely unbalanced, the entire tree structure will be rebuilt by Algorithm 1 with the current image of interest as the root, so as to achieve a more balanced tree structure.

Figure 7 shows the results by using our tree-transfer scheme to update the layout for the image collection shown in Figure 2. We also include the results of [13] and [22] in Figure 7 (c) which updates the layout by projecting a fix structure onto different 2-dimensional panels. From Figure 7, we can see that by our tree-transfer scheme, we can effectively update layouts according to the interested images of users while still keeping image correlations in the layout. Comparatively, although image correlations are also preserved in [13] and [22], its transferred layouts are obviously less appealing since these layouts are not adaptively optimized for different visualization panels.

## VI. EXPERIMENTAL RESULTS

Our algorithm is implemented by Java/Matlab and the experiments are performed on PC with 2.20GHz CPU and 4G RAM. We construct a dataset which includes 30 image collections and utilize these image collections to create 40 image visualization layouts (15 rectangular shaped layouts and 25 non-rectangular shaped layouts). The image collections are selected from various resources, including materials from a professional poster design dataset, searching results from Flickr, Google, and image retrieval systems like [49], and manually collected images. Some example image collections include the best-selling product pictures in Walmart, the world's top 50 company logos in Fortune, the vehicle pictures for Ford, the image search results of query "flower" returned by search engine, the mentor and player photos for the TV show "the Voice", and the images for Google zeitgeist 2012. These image collections cover a large variety of types and contents, as Figures 8-13 show. Besides, the total number of images in each image collection is between 30 and 100. Table 1 shows the image number distribution for these image collections.

Moreover, the default parameters of the TSVO scheme are set as: the maximum allowed overlap threshold $T_o = 0$ and the weights in Eqn. (8) are $\omega_t = \omega_l = 3\omega_s = 0.01\omega_o = 0.01\omega_{sp}$. These weights are selected from the experimental statistics which can properly balance the relative importance among different costs and create the satisfactory visualization results.

Table 1. The image number distribution for the 30 image collections

| Range of Image Number | 30-40 | 40-50 | 50-70 | 70-100 |
|---|---|---|---|---|
| Number of Image Collections | 6 | 10 | 8 | 6 |

● **Results of the tree construction scheme.** Figures 8-12 (a) show the image trees constructed by our PTC scheme over various image collections and user-defined properties (the properties are described in the figure captions). In particular, the lower figures in Figure 8 (a) and Figure 11 (a) show the image trees when the "Ford" car image collection is organized according to the "color" and the "vehicle type" properties, respectively. These figures further demonstrate the effectiveness of our PTC scheme in flexibly creating image trees for different user-defined properties.

● **Results of the two-step optimization scheme.** Figures 8-12 (b) show the final visualization results by our two-step visualization optimization (TSVO) scheme. In Figures 9 and 11, the rectangular shape is used while arbitrary shapes are applied in Figures 8, 10, and 12. From Figures 8-12 (b), we can see that our TSVO scheme can create satisfactory image collage results where (1): correlated images in the image trees in Figures 8-12 (a) are properly displayed close to each other, (2) the overlap among images are avoided, (3) the shape **SP** is fully utilized with few blank spaces inside **SP** and no images outside **SP**.

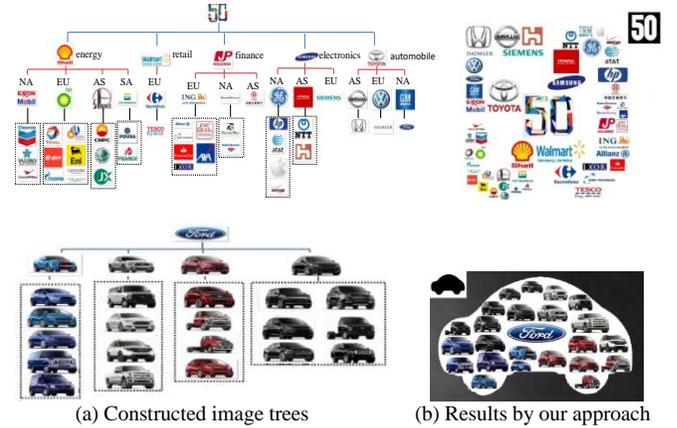

(a) Constructed image trees     (b) Results by our approach

Figure 8. Top: The collection world's 50 largest company logos organized by their "business category" property (e.g., electronics, energy), and their "region" property (e.g., Asian, North American, European). Down: The collection of Ford car images organized by "color" property.

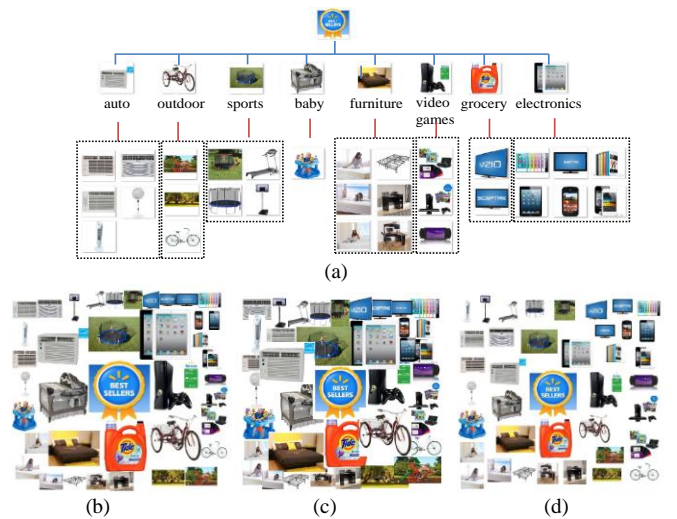

Figure 9. Results of our approach with different parameter values. (a) Image tree of the "the best-selling products in Walmart". (b) The visualization layout with default parameter values (the layout shape **SP** is a rectangle). (c) The result when increasing the overlap threshold $T_o = 20\%$. (d) The result when the image size upper bounds $w_i^u$ and $h_i^u$ in Eqn. (4) are decreased to $0.9 \cdot w_i^o$ and $0.9 \cdot h_i^o$.

● **Results of the tree transfer scheme.** Figure 10 shows the updated layouts by our tree transfer scheme. In Figure 10, (b) is the originally created visualization layouts by our approach while (c) and (d) are the updated layouts when different images of interest are selected. Figure 10 further demonstrates the effectiveness of our tree transfer scheme in creating suitable "personalized" layouts according to different user interested images.

● **Comparison with other methods.** Figure 11 compares the image visualization results of our approach with four existing algorithms under the rectangular layout shape: shape collage [9], Google's Picasa mosaic collage [12], Google's Picasa picture pile collage [12], and autocollage [3]. Also, Figure 12 compares our result with four methods under arbitrary layout shapes: hyperbolic project [13], shape collage [9], loupe collage [19], and packing layout [25]. Besides, in order to further analyze the difference between our approach and Centroidal-Voronoi-Tessellation (CVT)-based methods [25, 26], Figure 13 shows more comparison results between our approach and a CVT-based method (i.e., packing layout [25]). Note that in order to have a fair comparison, the results of the packing layout method [25] using the same input properties as our approach are also included (i.e., the result named "packing+our property" in Figure 13 (d)).

From Figures 11-13, we can see the following advantages of our algorithm compared with the other methods.

(1) Our approach can effectively display images with the desired property correlations under arbitrary shapes. Comparatively, in the compared methods [3, 9, 12, 19], images are disorderly placed in a layout. Besides, many methods [3, 12] are only limited to fixed layout shapes (e.g., rectangular) and cannot be extended to arbitrary shapes. Although the projection-based method [13] can embed the property correlations among images (Figure 12 (c)), it is unable to create arbitrary layout shapes.

(2) The overlap problems are effectively avoided in our approach. Comparatively, most compared algorithms [3, 9, 12, 13, 19] still include overlaps. Although the Picasa picture pile collage method [12] can effectively avoid the overlap, the grid-based layout strategy in this method is only suitable for rectangular shapes and has low flexibility to adapt with other layout shapes.

(3) Comparing the CTV-based method (i.e., packing layout [25]) and our approach, although both methods have the functionalities of creating arbitrary shapes and embedding image correlations, our approach still has obviously better results in the following parts: (i) The overlaps among images are effectively avoided by our approach while the overlap problem still exists in the results of the packing layout method [25] (i.e., the blue dotted circles in Figures 12-13). Specifically, in the top row of Figure 13, the blue-dotted circled images in (a) are severely overlapped by other images in the layout results in (c) and (d). (ii) The layout shape region is more effectively utilized by our approach. Comparatively, the layout results by [25] still include obvious blank regions, making images displayed with smaller sizes. (iii) Our approach can create more precise layouts according to user-desired properties [25]. For example, in Figure 13 (b) and (d), given the same input image properties, the packing layout method [25] cannot fully group images with similar properties (e.g., images in red solid circles in (a) are separately placed in (d)). This is because images with similar properties may be easily interrupted by other images during the layout optimization process due to the limited space in a shape region. However, by properly introducing and optimizing the "tree structure" and "property correlation" costs, our approach can still put together images with similar properties (e.g., images in red solid circles in Figure 13 (a) are arranged together in (b)).

● **Effectives of different parameter values.** Figure 9 shows the results of our approach with different parameter values where (b) is the result by default parameter values with a rectangular layout shape **SP**. (c) is the result when the overlap threshold $T_o$ is increased. From (c), it is clear that by tuning $T_o$, our approach can flexibly control the maximum allowed overlaps among images in the layout. Furthermore, Figure 9 (d) is the result when image size upper bounds $w_i^u$ and $h_i^u$ are decreased. We can see that due to the size limit, images in the layout are shrunken. This further demonstrates the effectiveness of our approach in flexibly controlling image sizes in a visualization layout.

● **Time statistics.** Table 2 shows the time statistics of different steps in our approach. Our approach is implemented by Matlab/Java. The time statistics is averaged over 40 image visualization results whose total image numbers are between 30 and 100. The image size ranges are 150×150-400×400. From Table 2, our approach takes about 25 seconds to create a final layout. Thus, the time complexity is suitable for the applications including poster design, collage creation, or album summarization. It should be noted that the processing time of our approach can be further improved by optimizing the code with C/C++, paralleling the process such as the local tuning step, and including GPU for speeding up the process.

Table 2. Timing statistics for our approach averaged over 40 visualization layouts with image number 30-100 and image size 150×150- 400×400 (note that the tree construction step includes the feature extraction process).

| Operation | Tree Construction | Global Optimization | Local Adjustment | Total |
|---|---|---|---|---|
| Time (s) | 9.77 | 8.57 | 7.29 | 25.63 |

● **User study.** Furthermore, Table 3 shows a user study test [6, 10] which evaluates the performance of different approaches. We asked 30 participants to score 40 sets of collages by different methods. The 40 collage sets include two groups where 15 collage sets belong to rectangular shaped group and 25 collage sets belong to non-rectangular shaped group. 15 shapes are used to create the 25 non-rectangular collage sets where each collage set includes collage results by different methods under the same shape.

The participants include 18 males and 12 females whose ages ranged from 20-60. Within these 30 participants, 10 of them have computer science-related background and the other

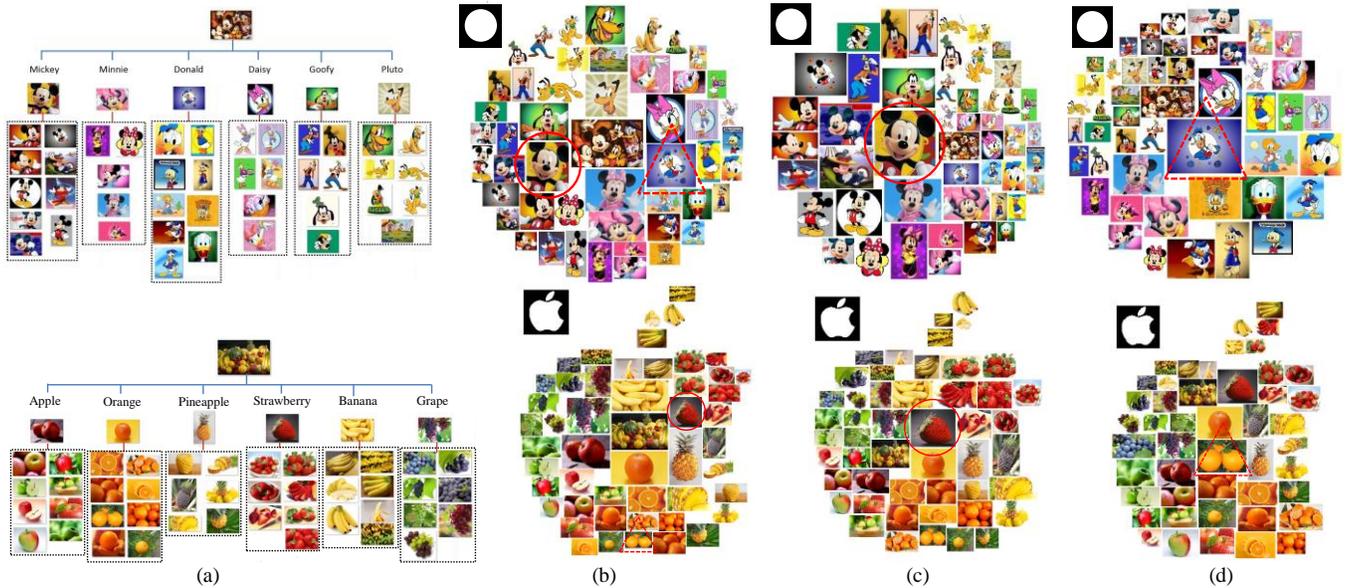

Figure 10. Top: The image collage of a photo album for "Mickey Mouse and Donald Duck". Down: The image collage of fruit images grouped by different kinds of fruits. (a): The constructed image trees; (b): The visualization layouts by our approach; (c): The updated layouts by the tree transfer scheme when the solid circle images are selected as the image of interest; (d): The updated layouts by the tree transfer scheme when the dashed triangle images are selected as the image of interest.

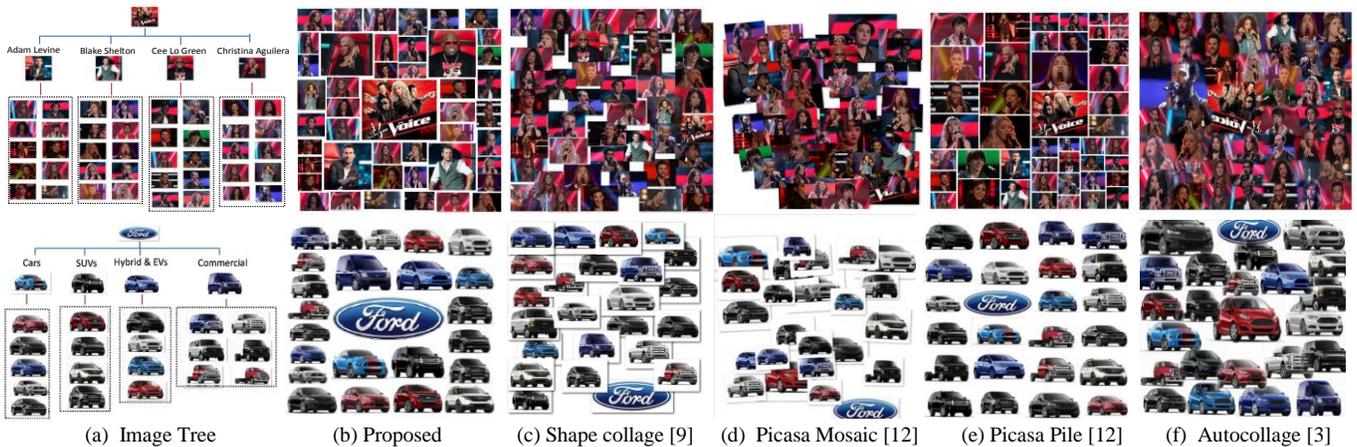

(a) Image Tree  (b) Proposed  (c) Shape collage [9]  (d) Picasa Mosaic [12]  (e) Picasa Pile [12]  (f) Autocollage [3]

Figure 11. Top: The collection of people (four mentors and top 40 artists) in the TV show the Voice (U.S. Season 3). The images with "mentor" properties are chosen as the nodes in the second level, and the artists are organized by "mentor" properties (i.e., the mentors they belong to). Down: The image collection of Ford vehicles' images. The images are organized based on the "vehicle type" property (i.e., Cars, SUVs, Hybrid & EVs, and Commercial).

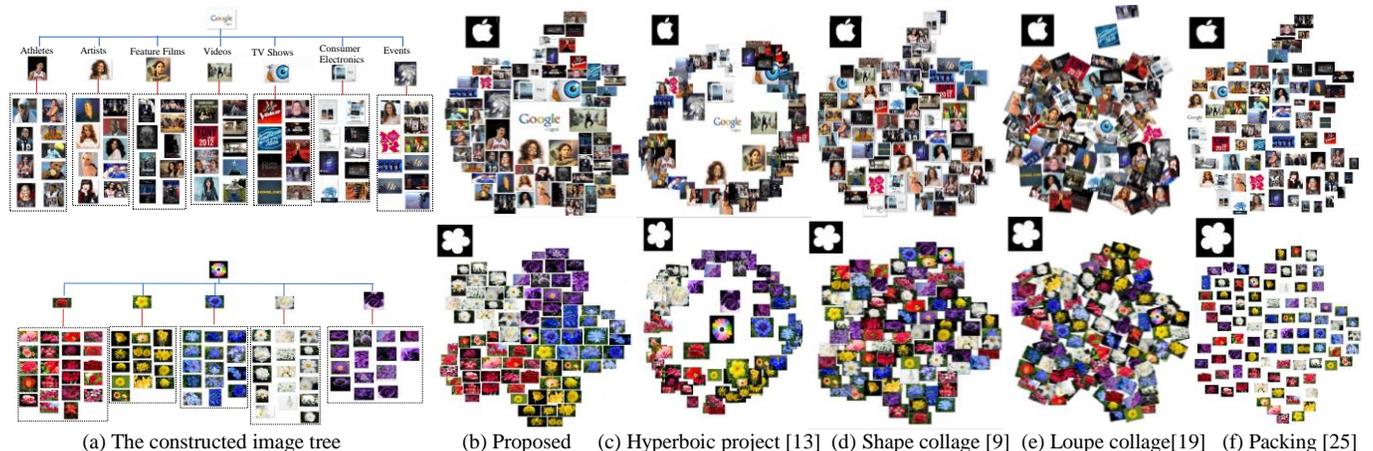

(a) The constructed image tree  (b) Proposed  (c) Hyperboic project [13]  (d) Shape collage [9]  (e) Loupe collage [19]  (f) Packing [25]

Figure 12. Top: The image collection for Google zeitgeist 2012 images (The list of 2012 Search Trends, i.e., the search queries with the highest amount of traffic over a sustained period in 2012 as compared to 2011). The images are organized based on the category properties (e.g., Athletes, Artists, Consumer Electronics, and Events). Down: The collection of an image search result with the query "flower". The images are organized by "color" properties. (Best viewed in color)

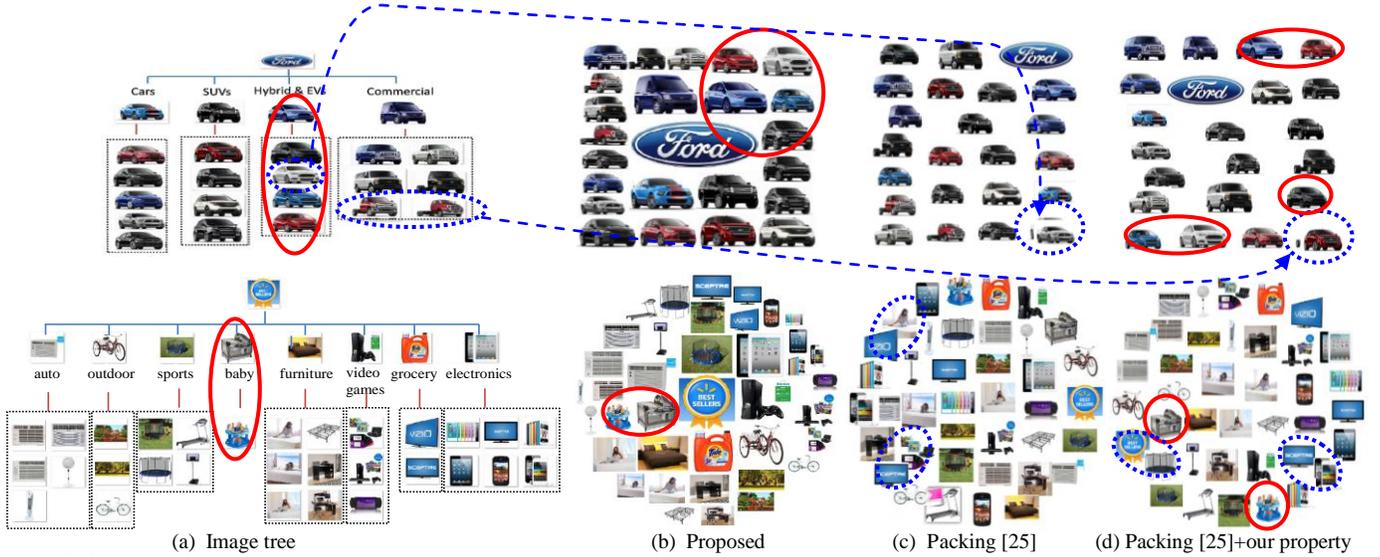

(a) Image tree     (b) Proposed     (c) Packing [25]     (d) Packing [25]+our property

Figure 13. Comparing the proposed approach with the packing-based method [25]: Images in red solid circles in (a) are arranged together by our approach in (b) but are placed separately in (d); Images in blue dotted circles in (c) and (d) are overlapped with each other. (best viewed in color)

Table 3. The average user score and confidence interval for different approaches
(× means the method is unable to create results under the required shape. The number before "±" is the average user score and the number after "±" is the 95% confidence interval)

| Shape \ Score \ Method | Criterion | **Our** | Shape Collage [9] | Mosaic Collage [12] | Picasa Pile Collage [12] | Auto Collage [3] | Loupe collage [19] | Hyperbolic project [13] |
|---|---|---|---|---|---|---|---|---|
| Rectangle Shape | Appealing | **8.3±0.45** | 7.0±0.53 | 5.7±0.65 | 7.6±0.38 | 6.7±0.31 | 6.9±0.31 | × |
| | Space | 8.6±0.33 | 8.1±0.43 | 4.4±0.53 | **8.8±0.26** | 7.1±0.48 | 7.4±0.39 | × |
| | Informative | **6.9±0.51** | 5.1±0.47 | 4.1±0.4 | 5.6±0.55 | 4.9±0.44 | 5.3±0.53 | × |
| Non-Rectangle Shape | Appealing | **8.6±0.31** | 7.7±0.28 | × | × | × | 7.5±0.43 | × |
| | Space | **8.9±0.29** | 8.1±0.41 | × | × | × | 8.0±0.36 | × |
| | Informative | **7.0±0.64** | 5.2±0.74 | × | × | × | 4.9±0.65 | × |

Table 4. Wilcoxon signed-rank test results for the user study scores
(× means the method is unable to create results under the required shape, "+" means our method has a higher and significantly different (at $p \leq 0.05$) score than the compared methods, '-' denotes that our method has a lower and significantly different (at $p \leq 0.05$) score, and "no sig. diff." means there is no significant difference (at $p \geq 0.05$) between the scores of the compared methods)

| Shape \ Score \ Method | Criterion | Our vs. Shape Collage [9] | Our vs. Mosaic Collage [12] | Our vs. Picasa Pile Collage [12] | Our vs. Auto Collage [3] | Our vs. Loupe collage [19] |
|---|---|---|---|---|---|---|
| Rectangle Shape | Appealing | + | + | + | + | + |
| | Space | + | + | no sig. diff. | + | + |
| | Informative | + | + | + | + | + |
| Non-Rectangle Shape | Appealing | + | × | × | × | + |
| | Space | + | × | × | × | + |
| | Informative | + | × | × | × | + |

20 do not have related backgrounds. Besides, 20 participants have undergraduate degrees and the other 10 have graduate degrees. All the subjects are without visual problems.

In this user study, three questions are asked to users: "Are the collages visually appealing and satisfying (*appealing*)", "Does the collage make good use of the space inside a shape and properly avoid overlaps? (*space*)", and "Is the collage an informative summarization which keeps the correlations among images? (*correlation*)". Basically, the "appealing" question reflects the overall quality of a collage and questions of "space" and "correlation" measure the quality of a collage from two important aspects.

User are required to give a score to each of the questions ranged from excellent (10) to poor (0). In order to avoid evaluation biases, the method information of image collages is concealed from the participants and collages from different methods are randomly placed. The scores for each method are averaged over 30 participants and over all its collage results in the rectangular shape group and the arbitrary shape group, respectively.

Table 3 shows the average user scores for different methods together with the 95% confidence interval [43]. Besides, Table 4 shows the Wilcoxon signed-rank test results [42] by comparing the user scores of our approach with each of the compared method. Table 4 indicates the existence of significant user-score differences between our approach and the compared methods.

From Tables 3 and 4, we can further draw the following observations:

(1) Except for the "space" score under rectangular shape, our approach has the highest scores for all the three questions in both the rectangular shape and the non-rectangular shape groups. This further demonstrates that our approach can create more appealing and better-organized visualization results than the compared methods.

(2) In the rectangular shape group, our approach and the Picasa pile collage method have higher "appealing" scores than the other methods. Since the major difference between "our approach/the Picasa pile collage method" and the other methods is the avoidance of image overlaps (as shown in Fig. 11), this also demonstrates that overlap is an important issue and unsuitable overlaps will obviously affect the visual satisfaction of collages.

(3) Furthermore, although our approach has lower "space" score than the Picasa pile collage method under the rectangular shape group (Table 3), this score difference is not significant as in shown Table 4. This implies that both methods have similar performances in controlling space and overlap when a rectangular shape is utilized. However, note that our approach differs from the Picasa pile collage method in that (a) Our approach can create collages under various shapes while the Picasa pile collage method is only limited to rectangular shapes; (b) Our approach can properly arrange images according to images' correlations in a layout. And the higher "appealing" score of our approach further demonstrates that proper image arrangements can effectively increase the degree of the visual satisfaction.

## VII. Conclusion

In this paper, we propose an approach which can handle both issues of visualizing image collections into arbitrary layout shapes and arranging images according to user-defined semantic or visual correlations.

Future works will include: (1) allowing images to rotate in the layout; (2) combining with image-resizing algorithms (e.g., image re-targeting) to more efficiently model the size change of images; (3) more precise ways to extracting the visual or semantic features from images; (4) optimizing our approach for speeding up the processing time.

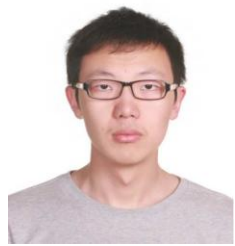

**Xintong Han** is currently a PhD student at University of Maryland, College Park, USA. He received his B.S. degree in electrical engineering. from Shanghai Jiao Tong University, China, in 2013.

His research interests are computer vision, machine learning, and multimedia.

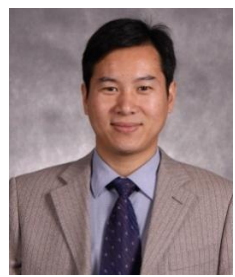

**Chongyang Zhang (M'2012),** received the Ph.D. degree from Shanghai Jiao Tong University, Shanghai, China, in 2008. He is currently an Associate Professor at the Department of Electronic Engineering, Shanghai Jiao Tong University, Shanghai China.

Dr. Zhang's research interests are in the area of video processing theory and application, in particular, video coding, video communications, intelligent video surveillance, and embedded multimedia system. He has published over 20 international journal/conference papers on these topics. As the principal investigator, he has owned or joint over 10 projects or grants from various agencies in China, such as the National Natural Science Foundation of China (NSFC) project, National Key Technology R&D Program of the 12th Five-Year-Plan of China. He has served on various IEEE conferences or other journals as a reviewer. Also, he acts as a member of IEEE and IEICE.

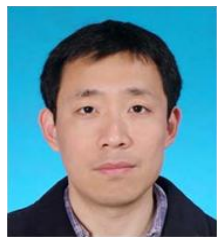

**Weiyao Lin** received the B.E. degree from Shanghai Jiao Tong University, China, in 2003, the M.E. degree from Shanghai Jiao Tong University, China, in 2005, and the Ph.D degree from the University of Washington, Seattle, USA, in 2010, all in electrical engineering. Currently, he is an associate professor at the Department of Electronic Engineering, Shanghai Jiao Tong University.

His research interests include video processing, machine learning, computer vision and video coding & compression.

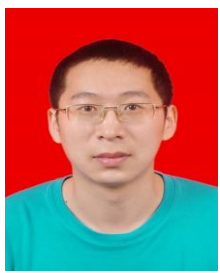

**Mingliang Xu** received the B.E. and M.E. degree from Zhengzhou University, China, in 2005 and 2008, respectively, and the Ph.D degree from the State Key Lab of CAD&CG at Zhejiang University, China, in 2012, all in computer science. Currently, he is an associate professor in the School of Information Engineering of Zhengzhou University, China, and also is the director of CIISR ( Center for Interdisciplinary Information Science Research). His research interests include computer graphics and computer vision.

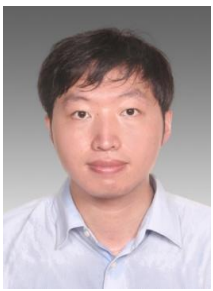

**Bin Sheng** received his BA degree in English and BE degree in Computer Science from Huazhong University of Science and Technology in 2004, MS degree in software engineering from the University of Macau in 2007, and PhD degree in Computer Science from The Chinese University of Hong Kong in 2011. He is currently an assistant professor in the Department of Computer Science and Engineering at Shanghai Jiao Tong University. His research interests include virtual reality, computer graphics, and image based techniques.

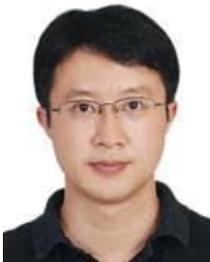

**Tao Mei (M'07-SM'11)** is a Lead Researcher with Microsoft Research, Beijing, China. He received the B.E. degree in automation and the Ph.D. degree in pattern recognition and intelligent systems from the University of Science and Technology of China, Hefei, China, in 2001 and 2006, respectively. His current research interests include multimedia information retrieval and computer vision. He has authored or co-authored over 100 papers in journals and conferences, 10 book chapters, and edited three books. He holds 13 U.S. granted patents and more than 20 in pending. Tao was the recipient of several paper awards from prestigious multimedia journals and conferences, including the IEEE Circuits and Systems Society Circuits and Systems for Video Technology Best Paper Award in 2014, the IEEE Trans. on Multimedia Prize Paper Award in 2013, and the Best Paper Awards at ACM Multimedia in 2009 and 2007, etc. He is an Associate Editor of IEEE Trans. on Multimedia, ACM/Springer Multimedia, and Neurocomputing, a Guest Editor of six international journals. He is the General Co-chair of ACM ICIMCS 2013, the Program Co-chair of IEEE ICME 2015, IEEE MMSP 2015 and MMM 2013. He is a Senior Member of the IEEE and the ACM.